\documentclass[aps,prd,preprintnumbers,groupedaddress,nofootinbib,amssymb,notitlepage,eqsecnum]{revtex4-2}
\usepackage{here}
\usepackage[dvipdfmx]{graphicx}
\usepackage{amsmath,amsthm,amssymb}
\usepackage{bm}
\usepackage{color}

\usepackage{amsfonts}
\usepackage{dcolumn}
\usepackage{hyperref}
\allowdisplaybreaks[1]
\usepackage{stackengine}

\newcommand{\be}{\begin{equation}}  
\newcommand{\ee}{\end{equation}}
\newcommand{\ba}{\begin{eqnarray}}
\newcommand{\ea}{\end{eqnarray}}

\newcommand{\rd}{{\rm d}}

\newcommand{\bem}{\begin{bmatrix}}
\newcommand{\eem}{\end{bmatrix}}

\allowdisplaybreaks

\begin{document}

\preprint{WUCG-25-01}

\title{Forecast constraints on the axion-photon 
coupling from interstellar medium heating}

\author{Makoto Amakawa$^{1}$}
\email{m.flygon.330@ruri.waseda.jp} 

\author{Tomohiro Fujita$^{2}$}
\email{fujita.tomohiro@ocha.ac.jp} 

\author{Shinji Tsujikawa$^{1}$}
\email{tsujikawa@waseda.jp} 

\affiliation{
$^1$Department of Physics, Waseda University, 
3-4-1 Okubo, Shinjuku, Tokyo 169-8555, Japan}

\affiliation{
$^2$Department of Physics, Ochanomizu University 2-1-1 Otsuka, 
Bunkyo, Tokyo 112-8610, Japan}

\begin{abstract}

In interstellar media characterized by a nonrelativistic plasma of electrons
and heavy ions, we study the effect of axion dark matter coupled to photons on 
the dynamics of an electric field. In particular, we assume the presence of 
a background magnetic field aligned in a specific direction.
We show that there is an energy transfer from the oscillating axion field to photons 
and then to the plasma induced by forced resonance.  
This resonance is most prominent for the axion 
mass $m_{\phi}$ equivalent to the plasma frequency $\omega_p$.  
Requiring that the heating rate of the interstellar 
medium caused by the energy transfer does not 
exceed the observed astrophysical cooling rate, 
we place forecast constraints on the axion-photon coupling $g$ 
for several different amplitudes of the background magnetic field $B_0$. 
By choosing a typical value $B_0=10^{-6}$~G, we find that, 
for the resonance mass $m_{\phi}=\omega_p$, 
the upper limit of $g$ can be stronger than those 
derived from other measurements in the literature.
With increased values of $B_0$, it is possible to
put more stringent constraints on $g$ for 
a wider range of the axion mass away from 
the resonance point.

\end{abstract}

\date{\today}


\maketitle

\section{Introduction}
\label{introsec}

The existence of a pseudo-Nambu Goldstone axion was originally introduced to 
address the strong CP problem 
in quantum chromodynamics (QCD) \cite{Peccei:1977hh,Weinberg:1977ma,Wilczek:1977pj}. 
The mass of QCD axions is related to the 
axion decay constant $f$, as 
$m_{\phi} \simeq 5.7 \times 10^{-6}~{\rm eV}\,
(10^{12}~{\rm GeV}/f)$ \cite{GrillidiCortona:2015jxo}.
The original model, which is 
not experimentally viable, was generalized to 
include heavy quarks \cite{Kim:1979if,Shifman:1979if} 
or an additional Higgs field \cite{Dine:1981rt,Zhitnitsky:1980tq}. 
The QCD axion in such extended 
versions can be light and stable for 
large $f$, being a good candidate for dark matter \cite{Preskill:1982cy,Abbott:1982af,Dine:1982ah,Dine:1982ah} 
(see Refs.~\cite{Kim:1986ax,Raffelt:1990yz,Kim:2008hd,Marsh:2015xka,DiLuzio:2020wdo,Kim:2014tfa} for reviews). 
In the context of string theory, ultralight axions can arise as Kaluza-Klein zero modes of 
anti-symmetric form fields \cite{Witten:1984dg,Svrcek:2006yi,Conlon:2006tq}. 
The mass of string axions can span over a vast range 
$10^{-33}\,{\rm eV} \lesssim  m_{\phi} \lesssim 10^{-10}$\,eV, depending on the compactification 
scheme \cite{Arvanitaki:2009fg,Acharya:2010zx,Cicoli:2012sz,Halverson:2017deq,Demirtas:2018akl}.

An axion $\phi$ may interact with photons 
through the Chern-Simons coupling
${\cal L}=-g \phi F_{\mu \nu} \widetilde{F}^{\mu \nu}/4$, where $g$ is a coupling constant and 
$F_{\mu \nu}$ is an electromagnetic field strength with $\widetilde{F}^{\mu \nu}$ its Hodge dual. 
In terms of the electric field ${\bm E}$ and the magnetic field ${\bm B}$, this interaction 
can be interpreted as the inner product $-g \phi {\bm E} \cdot {\bm B}$. 
Since the conversion of axions to photons can occur in the presence of 
external magnetic fields \cite{Sikivie:1983ip}, 
the laboratory experiments such as 
a ``light-shining-through-walls'' measurement \cite{ALPS:2009des} 
put upper limits on the coupling $g$. 
For the mass range 
$m_{\phi} \lesssim 10^{-4}$~eV, the ALPS 
collaboration has constrained 
$g \le 5 \times 10^{-8}$~GeV$^{-1}$ \cite{Ehret:2010mh}. 
By considering axions produced in the Sun that convert into X-rays 
within a laboratory magnetic field, the CAST experiment \cite{CAST:2017uph} 
has already placed a limit 
$g \le 0.66 \times 10^{-10}~\text{GeV}^{-1}$ 
for $m_{\phi} \le 2 \times 10^{-2}~\text{eV}$.
In contrast, the proposed IAXO experiment \cite{Armengaud:2014gea} 
has not obtained the data yet but is designed to increase the signal-to-noise ratio by approximately 4-5 orders of magnitude compared to CAST.

For light axions in the mass range 
$m_{\phi} \lesssim 10^{-10}$~eV, 
astrophysical observations 
put tightest bounds 
on the coupling constant $g$.
The interaction ${\cal L}=-g \phi F_{\mu \nu} \widetilde{F}^{\mu \nu}/4$ allows the 
generation of axions in the stellar plasma 
through a Primakoff process \cite{Primakoff:1951iae,Raffelt:1985nk}. 
Axions can eventually convert into gamma rays 
in the magnetic field of the Milky Way.
Since such gamma rays were not observed 
in the SN1987 event, this translates to the limit 
$g \le 5.3 \times 10^{-12}$~GeV$^{-1}$ for 
$m_{\phi} \le 4.4 \times 10^{-10}$~eV \cite{Payez:2014xsa}.
Similarly, axions generated inside stellar cores may convert into observable $X$-rays in the galactic 
magnetic field. The lack of observational evidence 
for $X$-rays from super star clusters leads to 
the limit $g \le 3.6 \times 10^{-12}$~GeV$^{-1}$ for 
$m_{\phi} \le 5 \times 10^{-11}$~eV \cite{Dessert:2020lil}. 
If we consider magnetic white dwarfs (MWDs), 
the axion-photon coupling generates photons polarized parallel to 
the direction of the magnetic field. Polarization measurements of thermal 
radiation from MWDs have placed the bound  
$g \le 5.4 \times 10^{-12}$~GeV$^{-1}$ for 
$m_{\phi} \le 3 \times 10^{-7}$~eV \cite{Dessert:2022yqq}. 
In the mass range of so-called fuzzy dark matter \cite{Hu:2000ke}, 
$m_{\phi} \sim 10^{-22}$~eV, 
axion searches by polarization plane rotation have been intensively 
studied~\cite{Fujita:2018zaj,POLARBEAR:2023ric,Xue:2024zjq}, and the limit 
due to the decrease in the cosmic microwave background (CMB) polarization is 
$g \lesssim 10^{-13}~{\rm GeV}^{-1}\,(m_{\phi}/10^{-22}~{\rm eV})$~\cite{Fedderke:2019ajk}.
We also note that cosmic birefringence inferred from 
observations of CMB \cite{Minami:2020odp,Diego-Palazuelos:2022dsq} 
can constrain $g$ for the even 
lighter mass range 
$m_{\phi} \lesssim 10^{-26}$~eV \cite{Carroll:1998zi,Lue:1998mq,Fujita:2020aqt,Fujita:2020ecn}.

In this paper, we will propose a new method of probing the axion-photon coupling 
through the observed cooling rate of interstellar media. We consider a nonrelativistic  
plasma of electrons and heavy ions with the electron number density $n_e$. 
In the absence of the axion-photon coupling, the collective movement of electrons 
gives rise to the plasma oscillation of an electric field ${\bm E}$ with the angular 
frequency $\omega_p=\sqrt{n_e e^2/m_e} \simeq 10^{-12}\,{\rm eV}\,
(n_e/10^{-3}~{\rm cm}^{-3})^{1/2}$, where $e$ is the elementary charge 
and $m_e$ is the electron mass \cite{Chen,Gibbon}. 
For the typical number density $n_e \approx 10^{-3}$~cm$^{-3}$ 
of dilute plasmas in the interstellar media, the plasma frequency 
$\omega_p$ is of order $10^{-12}$~eV. 
The existence of heavy ions works to generate 
friction for the dynamics of electrons characterized by the damping rate $\nu$. 
Since we typically have $\nu \ll \omega_p$ for the interstellar plasma, 
the electric field exhibits damped oscillations with the decay time scale 
$\nu^{-1}$, which is much longer than 
the oscillating time scale $\omega_p^{-1}$.
If there are background magnetic fields 
in the interstellar region, the interaction 
${\cal L}=-g \phi F_{\mu \nu} \widetilde{F}^{\mu \nu}/4$ 
can lead to the energy conversion from 
axions to photons in the plasma. 

A typical value of the galactic magnetic field in the Milky Way has been 
estimated to be of order $B_0=10^{-6}$~G \cite{Jansson:2012pc,Wadekar:2021qae}.
The observations of interstellar magnetic fields in the galactic center suggest that their 
amplitudes may span in the range $B_0=10^{-5}~{\rm G} \sim 10^{-3}$~G \cite{Ferriere:2009dh}. 
In this paper, we will consider an external magnetic 
field ${\bm B}_0$ aligned in a specific direction 
and compute the heating rate of interstellar media
induced by the axion-photon coupling. 
For this purpose, we deal with the axion as 
nonrelativistic dark matter oscillating around 
a minimum of the potential $V(\phi)=m_{\phi}^2 \phi^2/2$.
In the two-dimensional plane perpendicular to 
the direction of ${\bm B}_0$, the 
electric-field components still exhibit 
damped elliptic motions combined with 
the plasma oscillation and the cyclotron orbit in the normal manner. 
Along the direction of ${\bm B}_0$, however, we will show that the energy density 
of axions is converted to that of photons through forced resonance.
In particular, the energy transfer rate $\dot{Q}$ 
is largest for the resonance mass $m_{\phi}$ 
equivalent to the plasma frequency $\omega_p$.

The observations of interstellar media like Leo~T have constrained 
the cooling rate $\dot{C}$ and the temperature $T$ \cite{Matsuhara:1997zx,Lehner:2004ty,Wadekar:2019mpc,Wadekar:2022ymq}. 
They can be used to probe the properties of dark matter \cite{Chivukula:1989cc,Dubovsky:2015cca,Hardy:2016kme,Bhoonah:2018wmw,Bhoonah:2018gjb,Farrar:2019qrv,Bhoonah:2020dzs,Wadekar:2021qae,Shoji:2023pdg,Takeshita:2024ytb}. 
In our case, the heating rate $\dot{Q}$ 
induced by the axion-photon coupling should 
not exceed the observed cooling rate, which 
translates to the upper limit of $g$.
At the resonance point, we will derive an analytic formula 
for the bound on $g$. For the plasma frequency of order 
$10^{-12}~{\rm eV}$, the upper limit of $g$ 
at the resonance mass $m_{\phi}=\omega_p$ 
can be smaller than the MWD bound mentioned 
above by choosing the typical galactic magnetic field 
strength $B_0 \approx 10^{-6}$~eV. In particular, the constraint on $g$ 
at $m_{\phi}=\omega_p$ tends to be stronger 
for smaller $\nu$ due to the occurrence 
of sharper resonance. With the increase of $B_0$, 
we can place tighter bounds on $g$ for a broader 
mass range away from the resonance point. 
While the precise magnetic field strengths 
have been unknown for interstellar media with 
observed cooling rates, upcoming observations may be able to put interesting 
constraints on $g$ for the mass range  
$10^{-15}~{\rm eV} \lesssim m_{\phi} 
\lesssim 10^{-9}~{\rm eV}$.

This paper is organized as follows. 
In Sec.~\ref{plasmasec}, we revisit 
the fundamental aspect of nonrelativistic plasmas and present observational constraints on the cooling rates of interstellar media known in the literature.
In Sec.~\ref{dmsec}, we consider the 
axion-photon coupling on the background of
an external magnetic field in the plasma and derive solutions to the electric field under the approximation that 
spatial gradient terms are neglected 
relative to time-dependent terms.
In Sec.~\ref{heatsec}, we compute the heating rate $\dot{Q}$ of interstellar media induced by forced resonance and put upper bounds on $g$ derived from the condition 
$\dot{Q} \le \dot{C}$. 
Sec.~\ref{consec} is devoted to conclusions.
We use a natural unit in which the speed of light $c$, the reduced Planck constant $\hbar$, and the Boltzmann constant $k_B$ are 1. 
We adopt the metric signature $(-,+,+,+)$.

\section{Plasma in interstellar media}
\label{plasmasec}

We consider an interstellar medium modeled by a 
nonrelativistic plasma of electrons and heavy ions 
with a thermal temperature $T$ \cite{Chen,Gibbon}.
As in the standard plasma, the total charge 
of the system is assumed to be zero. Since the mass of ions is much larger 
than the electron mass, we ignore the ion velocity ${\bm u}_{i}$ relative to the electron velocity ${\bm u}_{e}$. 
Then, the total current is approximately 
given by 
\be
J^{\mu}=\left( 0, {\bm J} \right)=
\left( 0, -n_e e\, {\bm u}_{e} \right)\,. 
\label{Jmu}
\ee
We deal with the electron (mass $m_e$ and charge $-e$) as a nonrelativistic particle. 
We also neglect general relativistic effects 
on the dynamics of electrons and electromagnetic fields.
In the presence of an electric field 
${\bm E}$ and a magnetic field ${\bm B}$, the Newtonian equation of motion for 
the electron yields
\be
m_e \dot{{\bm u}}_e+m_e \left( {\bm u}_e \cdot \nabla \right) 
{\bm u}_e=-e \left( {\bm E}+{\bm u}_e \times {\bm B} 
\right)-m_e \nu {\bm u}_e\,,
\label{ele}
\ee
where a dot represents the derivative 
with respect to 
time $t$, $\nabla \equiv (\partial_1, \partial_2, \partial_3)$ 
with $\partial_i \equiv \partial/\partial x^i$ ($x^i$ is the spatial position), 
and $\nu$ is a friction constant between electrons and ions. 
We can express $\nu$ in the form \cite{Dubovsky:2015cca}
\be
\nu=\frac{4\sqrt{2\pi} \alpha^2 n_e}{6(m_e T^3)^{1/2}}
\ln \Lambda_c
\simeq 1.2 \times 10^{-24}~{\rm eV} \left( 
\frac{n_e}{10^{-3}~{\rm cm}^{-3}} 
\right) \left( \frac{10^4~{\rm K}}{T} \right)^{3/2} 
\ln \Lambda_c\,,
\label{nu}
\ee
where $\alpha=e^2/(4\pi)=1/137.036$ is the fine structure constant, and 
\be
\Lambda_c=\frac{4\pi T^3}{\alpha^3 n_e} \simeq
2.7 \times 10^{24} \left( \frac{10^{-3}~{\rm cm}^{-3}}{n_e} \right) 
\left( \frac{T}{10^4~{\rm K}} \right)^3\,.
\label{Lambdac}
\ee

Since we are considering nonrelativistic electrons ($|{\bm u}_e| \ll 1$), the 
contribution $m_e \left( {\bm u}_e \cdot \nabla \right) {\bm u}_e$ to Eq.~(\ref{ele}) 
can be neglected relative to the linear 
term in $ {\bm u}_e$.
Eq.~(\ref{ele}) is approximately given by 
\be
m_e \dot{{\bm u}}_e \simeq 
-e \left( {\bm E}+{\bm u}_e \times {\bm B} 
\right)-m_e \nu {\bm u}_e\,.
\label{ele2}
\ee
Assuming that $n_e$ is constant, 
the current ${\bm J}=-n_e e\, {\bm u}_{e}$ obeys 
\be
\dot{{\bm J}} \simeq \omega_p^2 {\bm E}
-\frac{e}{m_e}{\bm J} \times {\bm B}
-\nu {\bm J}\,,
\label{Jeq}
\ee
where 
\be
\omega_p=\sqrt{\frac{n_e e^2}{m_e}}
\simeq 1.2 \times 10^{-12}~{\rm eV} 
\left( \frac{n_e}{10^{-3}~{\rm cm}^{-3}} 
\right)^{1/2}
\label{wp}
\ee
is the plasma frequency. 
A dwarf galaxy Leo~T has the electron number density  
$n_e \approx 10^{-3}$~cm$^{-3}$ in the warm neutral region \cite{Wadekar:2022ymq}, 
in which case $\omega_p$ is of order $10^{-12}$~eV. 
In the Milky Way, there is a model for the free electron density 
called NE2001, which indicates that $n_e$ spans in the range 
$10^{-4}~{\rm cm}^{-3} \lesssim n_e \lesssim 0.1~{\rm cm}^{-3}$ depending 
on the region inside the galaxy \cite{Cordes:2002wz}. 
In the vicinity of the Solar system, the electron number density is 
$n_e \approx 0.04$~cm$^{-3}$, whose value is chosen to estimate 
$\omega_p$ in Ref.~\cite{Dubovsky:2015cca}. 
In this paper, we choose $n_e=10^{-3}$~cm$^{-3}$ 
as the typical electron number density.

In Table~\ref{tabinter}, we summarise the observed gas-rich 
astrophysical media, their temperatures, and the astrophysical 
cooling rate of the gas $\dot{C}$ for each medium. 
The electron-ion friction constant $\nu$ can be computed according to 
the relation (\ref{nu}) with (\ref{Lambdac}). 
For $n_e=10^{-3}$~cm$^{-3}$, we have $\nu \simeq 1.4 \times 
10^{-22}$~eV for WNM (Leo~T), 
which is much smaller than $\omega_p \simeq 1.2 \times 10^{-12}$~eV.
The property $\nu \ll \omega_p$ also holds for the other 
interstellar media shown in Table~\ref{tabinter}.
By using the observed values of $\dot{C}$, it is possible to put constraints 
on the coupling $g$ between axions and electromagnetic fields, 
which will be addressed in the subsequent sections.

\begin{table}
\begin{tabular}{|c|c|c|}
\hline
Interstellar media&  $T$~(K)  & $\dot{C}$~(erg\,cm$^{-3}$\,s$^{-1}$) \\
\hline  
WNM (Leo T)  & 6100 & $7 \times 10^{-30}$\\
\hline  
CNM (G33.4$-$8.0) &  400 & $1.46 \times 10^{-27}$\\
\hline  
MC (MW)  &  50 & $3.16 \times 10^{-24}$\\
\hline  
CNM (MW)  & 100 &  $2.02 \times 10^{-24}$\\
\hline  
WNM (MW)  &  5000 &  $8.3 \times 10^{-27}$ \\
\hline  
WIM (MW)  & $10^4$ & $5.4 \times 10^{-26}$\\
\hline  
HIM (MW)  & $10^6$ & $1.3 \times 10^{-27}$\\
\hline  
\end{tabular}
\caption{\label{tabinter}
Interstellar media, their temperatures $T$ and cooling rates $\dot{C}$. 
The different phases are labeled as MC (molecular clouds), CNM (cold neutral medium), 
WNM (warm neutral medium), WIM (warm ionized medium), and HIM (hot ionized medium). 
MW denotes the data in the Milky Way and Leo T is a gas-rich dwarf galaxy. 
We take the observed values of $T$ and $\dot{C}$ from 
Ref.~\cite{Wadekar:2022ymq}. 
}
\end{table}

\section{Axions coupled to electromagnetic fields}
\label{dmsec}

As a candidate for dark matter, we consider 
a pseudo-scalar axion field $\phi$ with 
a constant mass $m_{\phi}$. 
The axion potential is given by 
$V(\phi)=m_{\phi}^2 \phi^2/2$, which follows 
from $V(\phi)=m_\phi^2 f^2 [1-\cos(\phi/f)]$ 
as a leading-order term by the expansion around 
$\phi=0$. The axion is coupled to electromagnetic 
fields with the interacting Lagrangian 
$-g \phi F_{\mu \nu} \widetilde{F}^{\mu \nu}/4$, where $g$ is the coupling constant,
$F_{\mu \nu}=\partial_{\mu}A_{\nu}-\partial_{\nu}A_{\mu}$ 
is the electromagnetic field strength of 
a four-dimensional gauge field $A_{\mu}$,  
and $\widetilde{F}^{\mu \nu}=F_{\alpha \beta} {\cal E}^{\alpha \beta \mu \nu}/2$ is the dual tensor of $F_{\alpha \beta}$. 
Note that ${\cal E}^{\alpha \beta \mu \nu}$ is an anti-symmetric Levi-Civita tensor 
with the component 
${\cal E}^{0123}=
-1/\sqrt{-\bar{g}}$, where $\bar{g}$ 
is a determinant of the background 
metric tensor $\bar{g}_{\mu \nu}$.

The four current (\ref{Jmu}) in the plasma is coupled to $A_{\mu}$ through the interacting Lagrangian $J^{\mu}A_{\mu}$. 
Since we consider the weak gravitational 
regime in which general relativistic effects 
on the dynamics of electrons 
and electromagnetic fields can be 
neglected, the background spacetime 
is approximated by the Minkowski line element, $\rd s^2=-\rd t^2+\delta_{ij} 
\rd x^i \rd x^j$, so that $-\bar{g}=1$.
The action of such a system is given by 
\be
{\cal S}=\int {\rm d}^4 x \left[ 
-\frac{1}{2} \partial_{\mu} \phi
\partial^{\mu} \phi-\frac{1}{2}m_{\phi}^2 \phi^2 
-\frac{1}{4}F_{\mu \nu}F^{\mu \nu}
+J^{\mu} A_{\mu} 
-\frac{1}{4}g \phi F_{\mu \nu} \widetilde{F}^{\mu \nu}
\right]\,.
\label{action}
\ee
Varying (\ref{action}) with respect 
to $\phi$ and $A_{\mu}$,  
respectively, we obtain 
\ba
& &
\ddot{\phi}-\nabla^2 \phi+m_{\phi}^2 \phi
+\frac{1}{4}gF_{\mu \nu} \widetilde{F}^{\mu \nu}=0\,,
\label{phi} \\
& &
\square A_{\nu}-\partial_{\nu}\partial^{\mu}A_{\mu}
+g\,\partial^{\mu}\phi \widetilde{F}_{\mu \nu}
+J_{\nu}=0\,,
\label{Aeq}
\ea
where $\square \equiv \partial^{\mu}
\partial_{\mu}$.
The nonvanishing components of $F_{\mu \nu}$ are 
$F_{i0}=-F_{0i}=E_i$ (with $i=1,2,3$) and 
$F_{12}=-F_{21}=B_3$, $F_{23}=-F_{32}=B_1$, 
$F_{31}=-F_{13}=B_2$, where ${\bm E}=(E_1, E_2, E_3)$ 
and ${\bm B}=(B_1, B_2, B_3)$ are 
the electric and magnetic fields, respectively. 
We can express the relation 
$F_{\mu \nu}=\partial_{\mu}A_{\nu}-\partial_{\nu}A_{\mu}$ 
in the following form 
\be
{\bm E}=-\dot{\bm A}+\nabla A_0\,,\qquad 
{\bm B}=\nabla \times {\bm A}\,.
\label{EB}
\ee
Since we are considering the current vector 
$J_{\mu}=\left( 0, {\bm J} \right)$ in the plasma, we can express Eq.~(\ref{phi}) and $\nu=0, i$ components of Eq.~(\ref{Aeq}), as
\ba
& &
\ddot{\phi}-\nabla^2 \phi+m_{\phi}^2 \phi
+g {\bm E} \cdot {\bm B}=0\,,
\label{beq1}\\
& &
\ddot{A}_0-\nabla^2 A_0+\partial_0 (\partial^{\mu}A_{\mu})
+g \nabla \phi \cdot {\bm B}=0\,,
\label{beq2}\\
& &
\ddot{{\bm A}}-\nabla^2{\bm A}
+\nabla \left( \partial^{\mu} A_{\mu} \right)
+g ( \dot{\phi} {\bm B} +\nabla \phi \times {\bm E})
-{\bm J}=0\,,
\label{beq3}
\ea
where ${\bm A}=(A_1,A_2,A_3)$.

Under the gauge transformation $\tilde{A}_{\mu}=A_{\mu}
-\partial_{\mu}\chi$, the invariance of the action (\ref{action}) 
is ensured under the current conservation 
$\partial_{\mu} J^{\mu}=0$. 
From Eq.~(\ref{Jmu}), this translates to the condition $\nabla \cdot {\bm u}_e=0$, implying that our plasma fluid is incompressible. 
The residual gauge degree of freedom 
can be fixed by choosing the Lorentz gauge condition $\partial^{\mu} \tilde{A}_{\mu}=0$, 
so that $\chi$ obeys 
$\square \chi=\partial^{\mu}A_{\mu}$. 
In the following, we choose the gauge 
$\partial^{\mu} \tilde{A}_{\mu}=0$
and omit the tilde from the transformed fields, 
under which the two terms $\partial_0 (\partial^{\mu}A_{\mu})$ and 
$\nabla \left( \partial^{\mu} A_{\mu} \right)$ in 
Eqs.~(\ref{beq2}) and (\ref{beq3}) vanish.

In this paper, we focus on the homogeneous part 
of the axion field $\phi$.
We decompose $\phi$ into the homogeneous background part and the perturbed part, as 
\be
\phi=\phi_0(t)+\delta \phi (t, {\bm x})\,,
\ee
where $\phi_0$ is a function of $t$ alone, 
while $\delta \phi$ depends 
both on $t$ and the spatial 
position ${\bm x}$.
We assume that the latter is negligible 
relative to the former,
\be
|\delta \phi (t, {\bm x})| \ll |\phi_0(t)|\,.
\ee
So long as the inequality 
\be
|g {\bm E} \cdot {\bm B}| \ll m_{\phi}^2 \bar{\phi}_0
\label{apro1}
\ee
holds in Eq.~(\ref{beq1}), 
where $\bar{\phi}_0$ is the amplitude 
of $\phi_0$, the homogeneous mode 
of $\phi$ obeys 
\be
\ddot{\phi}_0+m_{\phi}^2 \phi_0 \simeq 0\,.
\ee
This is integrated to give 
\be
\phi_0(t)=\bar{\phi}_0 \cos(m_{\phi} t)\,,
\label{phiba}
\ee
where the phase at $t=0$ is set to 0 upon the choice of suitable initial conditions. 
When the background time-dependent 
part $\phi_0(t)$ oscillates around the potential minimum, it is known that 
axions act as cold dark matter. 
As long as the condition (\ref{apro1}) is satisfied, the generation of the perturbed part $\delta\phi$ from $\phi_0$ can be ignored.
 In Appendix~I, we will confirm the validity of the approximation (\ref{apro1}).

For the gauge field $A_{\mu}$, we perform the following Fourier transformation 
\be
A_{\mu}(t,{\bm x})=\int \frac{{\rm d}^3 k}{(2\pi)^3}\, e^{i {\bm k}\cdot {\bm x}}
\tilde{A}_{\mu} (t,{\bm k}) \,,
\ee
where ${\bm k}$ is a wavenumber with 
$k=|{\bm k}|$. In the following, we will 
omit the tilde for the quantities in Fourier space. We are interested in the regime where 
the spatial gradient term 
$|\nabla \delta \phi|$ is much smaller than $|\dot \phi_0|$, i.e., 
$|\nabla \delta \phi| \ll |\dot \phi_0|$, 
which holds for $k \lesssim m_{\phi}$ 
and $|\delta \phi| \ll |\phi_0|$. 
Moreover, we focus on the case in which 
the inequality
\be
|\nabla \delta \phi \times {\bm E}| \ll |\dot{\phi} {\bm B}|
\label{phicon2}
\ee
is satisfied.
 As we see in Appendix~I, we consider the presence of a background magnetic field ${\bm B}_0$ whose amplitude is not suppressed 
 compared to ${\bm E}$. 
In such a case, the condition (\ref{phicon2}) can be justified for 
$|\nabla \delta \phi| \ll |\dot \phi_0|$.
Eq.~(\ref{beq3}) approximately reduces to 
\be
\ddot{{\bm A}}+k^2{\bm A}
+g \dot{\phi}_0 {\bm B} 
-{\bm J}=0\,.
\label{master1}
\ee
This equation implies that, in the presence of the magnetic field $\bm B$, an oscillating axion $\dot{\phi}_0 \sim m_\phi \bar{\phi}_0$ gives a source term to $\bm A$ and the electromagnetic fields will be produced.
In contrast, the source term to $A_0$ is 
$g \nabla \phi\cdot \bm B$ in Eq.~\eqref{beq2}.
Since we have assumed 
$|\delta\phi|\ll |\phi_0|$, the produced $A_0$ is suppressed and its contribution to 
$\bm{E}=-\dot{\bm A}+\nabla A_0$ is subleading.
Furthermore, as we will see in the following, 
${\bm E}$ and ${\bm A}$ exhibit plasma oscillations 
with the approximate frequency $\omega_p$. 
In Fourier space, $\dot{\bm A}$ is of order $|\omega_p {\bm A}|$, whereas $\nabla A_0$ is 
of order $|k A_0|$. So long as 
\be
k \ll \omega_p\,,
\label{kcon}
\ee
and $|A_0|\lesssim |{\bm A}|$, 
we approximately find
\be
{\bm E} \simeq -\dot{\bm A}\,.
\label{EAre}
\ee
Then, the electron equation of 
motion (\ref{Jeq}) yields 
\be
\dot{{\bm J}} \simeq -\omega_p^2 \dot{{\bm A}}
-\frac{e}{m_e}{\bm J} \times {\bm B}
-\nu {\bm J}\,.
\label{master2}
\ee
For a given constant magnetic field 
${\bm B}={\bm B}_0$, 
the dynamics of ${\bm A}$ and ${\bm J}$ are 
known by 
solving the coupled differential Eqs.~(\ref{master1}) and (\ref{master2})
for ${\bm A}$ and ${\bm J}$. 

As a warm-up, let us revisit the case where 
${\bm B}={\bm 0}$. 
Under the condition (\ref{kcon}), we ignore 
the term $k^2 {\bm A}$ relative to 
$\ddot{\bm A}$ in Eq.~(\ref{master1}). 
On account of Eq.~(\ref{EAre}), we have 
${\bm J} \simeq -\dot{\bm E}$ and $\dot{\bm J} \simeq \omega_p^2 {\bm E}-\nu {\bm J}$ from 
Eqs.~(\ref{master1}) and (\ref{master2}), respectively. Combining these two equations 
leads to 
\be
\ddot{\bm E}+\nu \dot{{\bm E}}+\omega_p^2 {\bm E}=0\,, \qquad
{\rm for} \quad {\bm B}=0\,.
\label{Eeq0}
\ee
Provided that $\omega_p \gg \nu$, which holds for all the 
interstellar media shown 
in Table~\ref{tabinter}, the solution to 
Eq.~(\ref{Eeq0}) is given by 
\be
{\bm E} \simeq {\bm E}_0 e^{-\nu t/2} \cos (\omega_p t+\alpha_0)\,,
\label{Eso0}
\ee
where ${\bm E}_0$ is a constant vector, and $\alpha_0$ 
is an initial phase.
Thus, the electric field ${\bm E}$ and the current 
${\bm J} \simeq -\dot{\bm E}$ exhibit damped plasma 
oscillations with the frequency $\omega_p$. 
Since we are interested in the case $\omega_p \gg \nu$, 
the time scale of damping $t_d \approx 1/\nu$ induced by the friction is much longer than that of plasma oscillations $t_p \approx 1/\omega_p$, i.e., $t_d \gg t_p$. 

Now, let us consider the case in which there is a constant background magnetic field ${\bm B}_0$ along the $x$ direction in the Cartesian coordinate $(x,y,z)$, so that 
\be
{\bm B}_0=\left( B_0, 0, 0 \right)\,.
\ee
We note that the magnetic field has 
a contribution $\nabla \times {\bm A}$ 
besides the background value ${\bm B}_0$.
Since the former spatial derivative can be neglected in our approximation scheme, 
the magnetic field strength ${\bm B}$ 
is dealt as a time-independent constant (${\bm B} \simeq {\bm B}_0$). 
Expressing the components of ${\bm A}$ 
and ${\bm J}$ as 
${\bm A}=(A_x,A_y,A_z)$ and ${\bm J}=(J_x,J_y,J_z)$ 
in Eqs.~(\ref{master1}) and (\ref{master2}), we obtain 
\ba
& &
\ddot{A}_x+k^2 A_x-g \bar{\phi}_0 m_{\phi} B_0 \sin (m_\phi t) 
-J_x=0\,,\label{Axeq}\\
& &
\ddot{A}_y+k^2 A_y -J_y=0\,,
\label{Ayeq}\\
& &
\ddot{A}_z+k^2 A_z -J_z=0\,,
\label{Azeq}
\ea
and
\ba
& &
\dot{J}_x=-\omega_p^2 \dot{A}_x-\nu J_x\,,
\label{Jxeq}\\
& &
\dot{J}_y=-\omega_p^2 \dot{A}_y-\omega_c J_z
-\nu J_y\,,
\label{Jyeq}\\
& &
\dot{J}_z=-\omega_p^2 \dot{A}_z+\omega_c J_y
-\nu J_z\,,
\label{Jzeq}
\ea
where we used Eq.~(\ref{phiba}) and introduced 
\be
\omega_c \equiv \frac{eB_0}{m_e}\,.
\ee
We note that the quantity $\omega_c$, which corresponds to 
the cyclotron frequency, appears only for the dynamics in 
the $(y, z)$ plane. It is informative to express $\omega_c$ 
in the form 
\be
\omega_c=1.2 \times 10^{-14}~{\rm eV} 
\left( \frac{B_0}{10^{-6}~{\rm G}} \right)\,,
\ee
where we used the conversion of unit 
$1\,{\rm G}=1.95 \times 10^{-2}$~eV$^2$.
For a typical magnetic field strength $B_0 \approx 10^{-6}$~G, 
we have $\omega_c \simeq 10^{-14}$~eV. 
In such a case, so long as $n_{e}$ is not much less than the 
order $10^{-3}$~cm$^{-3}$, 
the plasma frequency (\ref{wp}) is much larger than 
$\omega_c$, i.e., $\omega_p \gg \omega_c$. 
For $B_0 \gtrsim 10^{-4}$~G, we have the 
opposite inequality $\omega_p \lesssim \omega_c$. 
Even for such large values of $B_0$, we are assuming the 
presence of a background magnetic field and an axion 
field which are both homogeneous, so that the local density 
of electrons is not significantly perturbed by them. 
Then, the approximation of an incompressible plasma 
for electrons ($\nabla \cdot {\bm u}_e=0$) does not lose 
its validity.

In Eqs.~(\ref{Axeq})-(\ref{Jzeq}), 
we observe that the dynamics of the system 
are separated into those along the $x$ direction and in the $(y,z)$ plane. 
In particular, the axion-photon 
coupling appears only for the dynamics 
in the $x$ direction.
In the $(y,z)$ plane, there is no energy transfer from axions to photons. 
Hence, we send the discussion of 
the $(y,z)$ plane to Appendix~II and concentrate on the $x$ direction 
in the main text.

Taking the time derivative of Eq.~(\ref{Axeq}) and using Eq.~(\ref{Jxeq}), 
we obtain the differential equation 
for $A_x$, as 
\be
\dddot{A}_x+\nu \ddot{A}_x+\left( k^2+\omega_p^2 \right) 
\dot{A}_x+\nu k^2 A_x=g \bar{\phi}_0 B_0 m_{\phi} 
\left[ m_{\phi} \cos (m_{\phi} t)+\nu \sin (m_{\phi} t) \right]\,.
\label{Axdif}
\ee
This has a special solution of the form 
\be
A_x^{(s)}={\cal D}_1 \cos (m_\phi t)+{\cal D}_2 \sin (m_\phi t)
={\cal D} \cos (m_\phi t -\alpha)\,,
\label{Axs}
\ee
where 
\ba
{\cal D}_1 &=& -\frac{\nu g \bar{\phi}_0 B_0 m_{\phi}^2 \omega_p^2}
{m_{\phi}^2 (m_{\phi}^2-\omega_p^2-k^2)^2+\nu^2 (m_{\phi}^2-k^2)^2}\,,
\label{D1}\\
{\cal D}_2 &=&\frac{g \bar{\phi}_0 B_0 m_{\phi} [m_{\phi}^2 \omega_p^2-(m_{\phi}^2+\nu^2)
(m_{\phi}^2-k^2)]}{m_{\phi}^2 (m_{\phi}^2-\omega_p^2-k^2)^2+\nu^2 (m_{\phi}^2-k^2)^2}\,,
\label{D2}
\ea
and 
\ba
{\cal D} &=& \sqrt{{\cal D}_1^2+{\cal D}_2^2}=
\frac{g \bar{\phi}_0 B_0 m_{\phi} \sqrt{m_{\phi}^2+\nu^2}}
{\sqrt{m_{\phi}^2 (m_{\phi}^2-\omega_p^2-k^2)^2+\nu^2 
(m_{\phi}^2-k^2)^2}}
\,,\\
\tan \alpha &=& \frac{{\cal D}_2}{{\cal D}_1}
=-\frac{m_{\phi}^2 \omega_p^2-(m_{\phi}^2+\nu^2)
(m_{\phi}^2-k^2)}
{\nu m_{\phi} \omega_p^2}\,.
\ea
The homogeneous solution to $A_x$ can be 
found by setting the right-hand side of 
Eq.~(\ref{Axdif}) to zero. 
Ignoring the $k$-dependent term on 
the left-hand side 
of Eq.~(\ref{Axdif}) under the condition (\ref{kcon}), we can derive the following 
homogeneous solution 
\be
A_x^{(h)} \simeq c_0+e^{-\nu t/2} \left[ c_1 \cos (\omega_p t)
+c_2 \sin (\omega_p t) \right]\,,
\label{Axh}
\ee
where $c_0$, $c_1$, and $c_2$ are integration constants.
Note that we have used the approximation $\omega_p \gg \nu$ 
for the derivation of Eq.~(\ref{Axh}).

The general solution to Eq.~(\ref{Axdif}) is the sum 
of Eqs.~(\ref{Axs}) and (\ref{Axh}), so that $A_x=A_x^{(s)}+A_x^{(h)}$. 
The amplitude of an oscillating mode in $A_x^{(h)}$ 
(with the frequency $\omega_p$) decays with the time 
scale of order $t_d=1/\nu$. 
As we see in Eqs.~(\ref{Axeq}) 
and (\ref{Jxeq}), the constant $c_0$ in $A_x^{(h)}$ does not affect the dynamics in the $x$ direction for $k \ll \omega_p$.
The special solution (\ref{Axs}), which oscillates with the frequency $m_{\phi}$, has a constant amplitude ${\cal D}$.
Then, the dominant contribution to $A_x$ arises from $A_x^{(s)}$. In particular, under the condition (\ref{kcon}), the largest contribution to ${\cal D}$ should come from the homogeneous mode corresponding to the 
$k \to 0 $ limit. 
Then, the resulting solution to $A_x$ 
is given by 
\be
A_x \simeq A_x^{(s)}
={\cal D}  \cos (m_\phi t -\alpha)\,,
\label{Axf}
\ee
where 
\be
{\cal D} \simeq \frac{g \bar{\phi}_0 B_0 \sqrt{m_{\phi}^2+\nu^2}}
{\sqrt{(m_{\phi}^2-\omega_p^2)^2+m_{\phi}^2 \nu^2}}\,,\qquad
\tan \alpha \simeq  
\frac{m_{\phi} (m_{\phi}^2-\omega_p^2+\nu^2)}{\nu \omega_p^2}\,.
\ee
We also obtain the electric field component 
\be
E_x=-\dot{A}_x \simeq m_{\phi}{\cal D} \sin (m_{\phi}t-\alpha)\,.
\label{Exf}
\ee
The amplitudes of $A_x$ and $E_x$ blow up around the axion mass 
$m_{\phi}=\omega_{p}$, which 
is a characteristic feature of 
forced resonance.
At $m_{\phi}=\omega_{p}$, the squared amplitude of $A_x$ has a peak value 
${\cal D}_{\rm M}^2=(g \bar{\phi}_0 B_0)^2 (\omega_p^2+\nu^2)/(\omega_p^2 \nu^2)$. 
The region in which ${\cal D}^2 \geq {\cal D}_{\rm M}^2/2$ corresponds to $\omega_{p}-\nu/2 \le m_{\phi} \le \omega_{p}+\nu/2$, 
so that the resonance width is $\Delta m_{\phi}=\nu$. Since we are now considering the case 
$\nu \ll \omega_p$, ${\cal D}^2$ has a sharp peak
around $m_{\phi}=\omega_p \pm \nu/2$.
The length (or time) scale associated with the resonance width 
$\Delta m_{\phi}=\nu$ corresponds to  $t_d=1/\nu$. 
Since $\nu \approx 10^{-22}$~eV for WNM (Leo~T), 
we have $t_d \approx 10^{15}~{\rm m} \approx 0.1$~pc.
On the other hand, the typical scale for the spatial variation of $n_e$ in the 
galactic plasma is of order $t_e \approx 0.1 \sim 1$~kpc, 
see e.g., Ref.~\cite{Cordes:2002wz}. 
Since $t_d \ll t_e$, the resonance position $m_{\phi}=\omega_{p}=\sqrt{n_e e^2/m_e}$ 
hardly changes during the time interval $t_d$. 
Thus, so long as $t_d \ll t_e$, the efficiency of resonance is not 
spoiled by the spatial variation of $n_e$.

\section{Heating rate induced by forced resonance}
\label{heatsec}

In Sec.~\ref{dmsec}, we have seen that 
the forced oscillation of $A_x$ induced by 
the axion coupled to electromagnetic fields 
does not lead to damping of the $E_x$ component. In this section, we compute the energy transferred from axions to photons. 

Taking the homogeneous limit $k \to 0$ in 
Eq.~(\ref{Axdif}) and 
using the property $E_x \simeq -\dot{A}_x$, 
we have 
\be
\ddot{E}_x+\nu \dot{E}_x-\omega_p^2 \dot{A}_x
=-g \bar{\phi}_0 B_0 m_{\phi} 
\left[ m_{\phi} \cos (m_{\phi} t)+\nu \sin (m_{\phi} t) \right]\,.
\ee
Integrating this equation with respect to $t$ leads to
\be
\dot{E}_x+\nu E_x-\omega_p^2 A_x=
-g \bar{\phi}_0 B_0 \left[ 
m_{\phi} \sin (m_{\phi} t) -\nu \cos (m_{\phi} t) \right]\,,
\label{Exre}
\ee
where the integration constant is set to 0 for consistency with 
the solution (\ref{Axf}). 
Multiplying Eq.~(\ref{Exre}) with 
$E_x=-\dot{A}_x$, 
we obtain the following relation 
\be
\frac{\rd}{\rd t} \left( \frac{1}{2}E_x^2
+\frac{1}{2}\omega_p^2 A_x^2\right)
=-g \bar{\phi}_0 B_0 \left[ 
m_{\phi} \sin (m_{\phi} t) -\nu \cos (m_{\phi} t) 
\right] E_x-\nu E_x^2\,.
\ee
We note that this has been derived by taking 
the large-scale limit $k \to 0$ in Fourier 
space, which amounts to neglecting all the 
spatial derivatives in real-space Eqs.~(\ref{beq1})-(\ref{beq3}). 
Since we are interested in the large-scale magnetic field
whose wavelength is of order 100~pc or more, this translates 
to the wavenumber $k \lesssim 10^{-25}$~eV. 
We are interested in the ranges of $\omega_p$ and $m_{\phi}$ 
spanning around $10^{-12}$~eV, in which case 
$k \ll \{ \omega_p, m_{\phi}\}$. In Eqs.~(\ref{D1}) and 
(\ref{D2}), the $k^2$ term is suppressed relative to 
$\omega_p^2$ and $m_{\phi}^2$ for the large-scale 
magnetic field of our interest. 
Hence it is a good approximation to take the limit $k \to 0$
for the estimation of $E_x$.
Then, we can interpret $\rho_x
=E_x^2/2 +\omega_p^2 A_x^2/2$ as 
the $x$-component of the photon energy density in real space obtained by keeping 
the time derivatives alone.
The temporal change of $\rho_x$ 
averaged over the oscillating period of the axion $T_{\phi}=2\pi/m_{\phi}$ is given by 
\be
\frac{1}{T_{\phi}} \int_0^{T_{\phi}} \frac{\rd}{\rd t} 
\left( \frac{1}{2}E_x^2 +\frac{1}{2}\omega_p^2 A_x^2\right) \rd t
=\dot{Q}-\dot{Q}_{\nu}\,,
\label{enere}
\ee
where 
\ba
\dot{Q} &=& -\frac{1}{T_{\phi}} \int_0^{T_{\phi}} g \bar{\phi}_0 B_0 \left[ 
m_{\phi} \sin (m_{\phi} t) -\nu \cos (m_{\phi} t) \right] E_x\,\rd t\,,\label{Q} \\
\dot{Q}_{\nu} &=& \frac{1}{T_{\phi}} \int_0^{T_{\phi}} \nu E_x^2\,\rd t\,.
\label{Qnu}
\ea
On using Eqs.~(\ref{Axf}) and (\ref{Exf}), 
we find that the left-hand side of 
Eq.~(\ref{enere}) vanishes. 
On the other hand, the energy transfer rate (\ref{Q}) arising from the coupling between axions and photons has a nonvanishing 
value $\dot{Q}=-(1/2)g \bar{\phi}_0 B_0 m_{\phi} {\cal D} (m_{\phi} \cos \alpha 
+\nu \sin \alpha)$. 
Exploiting the relations $\cos \alpha={\cal D}_1/{\cal D}$, 
$\sin \alpha={\cal D}_2/{\cal D}$ and taking the limit $k \to 0$ in 
Eqs.~(\ref{D1})-(\ref{D2}), we obtain
\be
\dot{Q}=\frac{g^2 \bar{\phi}_0^2 B_0^2 \nu m_{\phi}^2 (m_{\phi}^2+\nu^2)}
{2(m_{\phi}^2-\omega_p^2)^2+2m_{\phi}^2 \nu^2}\,.
\label{Qf}
\ee
The energy loss (\ref{Qnu}) arising from the friction term also has the nonvanishing value $\dot{Q}_{\nu}=(1/2)\nu m_{\phi}^2 {\cal D}^2$, i.e., 
\be
\dot{Q}_{\nu}=\frac{g^2 \bar{\phi}_0^2 B_0^2 \nu m_{\phi}^2 (m_{\phi}^2+\nu^2)}
{2(m_{\phi}^2-\omega_p^2)^2+2m_{\phi}^2 \nu^2}\,.
\label{Qnu2}
\ee
Thus, we find the following relation 
\be
\dot{Q}=\dot{Q}_{\nu}\,.
\label{Qba}
\ee
Due to this energy balance, 
the photon energy density $E_x^2/2 +\omega_p^2 A_x^2/2$ stays constant on the time average over oscillations.
This means that the energy transfer occurs from 
axions to photons with the heating rate given by Eq.~(\ref{Qf}) and that the injected energy density is lost by the plasma friction at the same rate~(\ref{Qnu2}). 
Thus the energy density flows sequentially from 
axions to photons, so that the plasma 
sustains its energy unlike the dynamics in the $(y,z)$ plane. 
For the axion mass $m_{\phi}=\omega_p$, 
the heating rate has a maximum value 
\be
\dot{Q}_{\rm max}=\frac{g^2 \bar{\phi}_0^2 B_0^2(\omega_p^2+\nu^2)}{2\nu}\,,
\qquad {\rm for}\quad
m_{\phi}=\omega_p\,.
\label{Qmax}
\ee
So long as the axion is the main 
source of dark matter, 
the corresponding energy density is 
estimated to be 
$\rho_{\rm DM}=\dot{\phi}^2/2+m_{\phi}^2 \phi^2/2
=m_{\phi}^2 \bar{\phi}_0^2/2$. 
For the plasma frequency in the 
range $\omega_p \gg \nu$, 
we can express $\dot{Q}_{\rm max}$, as 
\be
\dot{Q}_{\rm max} \simeq \frac{g^2 B_0^2 \rho_{\rm DM}}{\nu}\,.
\ee
Since $\dot{Q}_{\rm max}$  should not exceed the observed cooling rate $\dot{C}$ of interstellar media, we have the bound $\dot{Q}_{\rm max} \le \dot{C}$.\footnote{We note that this bound was also used for constraining the kinetic mixing between hidden and standard-model  
photons \cite{Dubovsky:2015cca}. 
In this case, the heating of interstellar 
media can occur through the resonant conversion from hidden photons to regular photons.} 
This translates to the upper limit 
on the coupling $g$, as
\be
g \le 1.9 \times 10^{-14}~{\rm GeV}^{-1} 
\left( \frac{B_0}{10^{-6}~{\rm G}} \right)^{-1}
\left( \frac{\nu}{10^{-22}~{\rm eV}} \right)^{1/2}
\left( \frac{\rho_{\rm DM}}{0.3~{\rm GeV}\,{\rm cm}^{-3}} \right)^{-1/2}
\left( \frac{\dot{C}}{10^{-27}~{\rm erg}~{\rm cm}^{-3}~{\rm s}^{-1}} \right)^{1/2}\,,
\label{gcon}
\ee
which is valid at the resonance point with 
the axion mass $m_{\phi}=\omega_{p}$.
For $B_0=10^{-6}$~G, $\nu=10^{-22}$~eV, 
$\rho_{\rm DM}=0.3$~GeV\,{\rm cm}$^{-3}$, 
and $\dot{C}=10^{-27}~{\rm erg}~{\rm cm}^{-3}~{\rm s}^{-1}$, 
we have the forecast constraint 
$g \le 1.9 \times 10^{-14}$~GeV$^{-1}$. 
This is tighter than the observational bound 
$g \le 5.4 \times 10^{-12}$~GeV$^{-1}$ 
extracted from polarization measurements 
of the MWD stars for the mass range 
$m_{\phi} \le 3 \times 10^{-7}$~eV \cite{Dessert:2022yqq}.
Should we obtain tighter observational 
bounds on the cooling rate $\dot{C}$, the constraint on $g$ 
becomes more stringent. 
For larger $B_0$, the upper limit on $g$ also 
decreases in proportion to $B_0^{-1}$.

\begin{figure}[ht]
\begin{center}
\includegraphics[height=3.5in,width=7.0in]{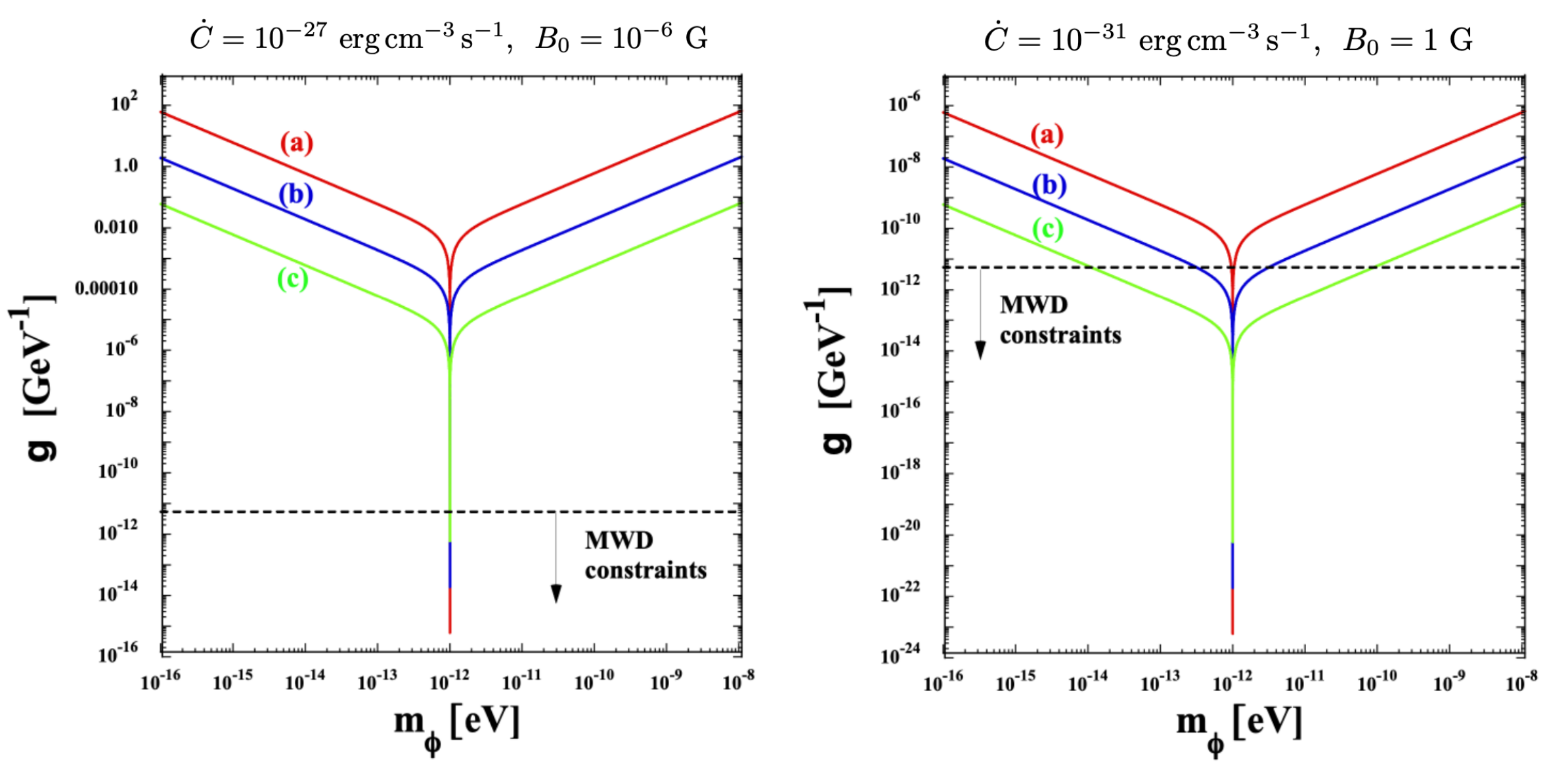}
\end{center}\vspace{-0.5cm}
\caption{\label{fig1}
The coupling constant $g$ versus the axion mass 
$m_{\phi}$ for $\dot{Q}$ equivalent to the two 
cooling rates $\dot{C}=10^{-27}$~erg\,cm$^{-3}$\,s$^{-1}$ 
(left) and $\dot{C}=10^{-31}$~erg\,cm$^{-3}$\,s$^{-1}$ (right). 
In the left and right panels, we choose the background 
magnetic field strength to be $B_0=10^{-6}$~G and 
$B_0=1$~G, respectively, with $\omega_p=
10^{-12}$~eV and $\rho_{\rm DM}=0.3\,{\rm GeV}$\,cm$^{-3}$ 
in both cases. The thick colored lines correspond to  
(a) $\nu=10^{-13}\omega_{p}$, (b) $\nu=10^{-10}\omega_{p}$, and 
(c) $\nu=10^{-7}\omega_{p}$, respectively. 
For decreasing $\nu$, the sharper resonance occurs with smaller minimum 
values of $g$. The dashed black lines represent the observational bound 
$g \le 5.4 \times 10^{-12}$~GeV$^{-1}$ derived from polarization measurements 
of the MWD stars for the axion mass range 
$m_{\phi} \le 3 \times 10^{-7}$~eV \cite{Dessert:2022yqq}.
}
\end{figure}

For the axion mass $m_{\phi}$ away from the plasma frequency $\omega_p$, 
the resonance does not occur efficiently. 
In this regime, the bound 
$\dot{Q} \le \dot{C}$ translates to 
$g \le g_{\rm max}$, where 
\be
g_{\rm max}=\frac{1}{B_0} \sqrt{\frac{\omega_p \dot{C}}{\rho_{\rm DM}}}
\sqrt{\frac{(r_{\phi}^2-1)^2+r_{\phi}^2 r_{\nu}^2}
{r_{\nu} (r_{\phi}^2+r_{\nu}^2)}}\,,
\label{gmax}
\ee
where 
\be
r_{\phi} \equiv \frac{m_{\phi}}{\omega_p}\,,\qquad 
r_{\nu} \equiv \frac{\nu}{\omega_p}\,.
\ee
Let us consider the case in which the inequalities 
$\nu \ll \omega_p$ and $\nu \ll m_{\phi}$ hold, 
i.e., $r_{\nu} \ll 1$ and $r_{\nu} \ll r_{\phi}$. 
At the resonance point ($r_{\phi}=1$) and the two asymptotic 
regimes $r_{\phi} \ll 1$ and $r_{\phi} \gg 1$, respectively, 
Eq.~(\ref{gmax}) has the following dependence 
\ba
g_{\rm max}& \simeq& \frac{1}{B_0} \sqrt{\frac{\nu \dot{C}}{\rho_{\rm DM}}}\,,
\qquad {\rm for} \quad 
m_{\phi}=\omega_p\,,\label{gm1}\\
g_{\rm max} & \simeq& \frac{\omega_p^2}{B_0 m_{\phi}}
\sqrt{\frac{\dot{C}}{\rho_{\rm DM} \nu}}\,,\qquad {\rm for} \quad 
m_{\phi} \ll \omega_p\,,\label{gm2}\\
g_{\rm max} & \simeq& \frac{m_{\phi}}{B_0}
\sqrt{\frac{\dot{C}}{\rho_{\rm DM} \nu}}\,,\qquad {\rm for} \quad 
m_{\phi} \gg \omega_p\,,\label{gm3}
\ea
where $g_{\rm max}$ in Eq.~(\ref{gm1}) coincides with 
the right-hand side of Eq.~(\ref{gcon}). 
In the light mass range $m_{\phi} \ll \omega_p$, the estimation 
(\ref{gm2}) shows that $g_{\rm max}$ increases with the decrease of $m_{\phi}$. 
In the heavy mass range $m_{\phi} \gg \omega_p$, from Eq.~(\ref{gm3}), 
$g_{\rm max}$ grows with the increase of $m_{\phi}$. 
In these two asymptotic regimes, 
for smaller $\nu$, 
we have larger values of $g_{\rm max}$. At the resonance point, 
$g_{\rm max}$ gets smaller for decreasing $\nu$. 
These properties are attributed to the fact that the resonance 
tends to be sharper for smaller $\nu$.

In Fig.~\ref{fig1}, we plot $g_{\rm max}$ versus $m_{\phi}$ for three different values 
of $\nu$. We choose two combinations of $\dot{C}$ and $B_0$ in the left and right 
panels. In both panels, the plasma frequency 
and the dark matter density are fixed to be  $\omega_p=10^{-12}$~eV and 
$\rho_{\rm DM}=0.3\,{\rm GeV}$\,cm$^{-3}$. 
For a given $\nu$, $g_{\rm max}$ has 
a minimum at the resonance mass $m_{\phi}=\omega_p$. For decreasing $\nu$, 
the minimum values of 
$g_{\rm max}$ get smaller.
In both the two asymptotic mass 
regions $m_{\phi} \ll \omega_{p}$ and $m_{\phi} \gg \omega_{p}$, 
we can confirm that $g_{\rm max}$ tends to 
increase for smaller $\nu$.

Since the ratio $r_{\nu}=\nu/\omega_p$ of interstellar media 
given in Table~\ref{tabinter} is in the region $10^{-13} \lesssim r_{\nu} \lesssim 10^{-7}$, 
we have chosen three different values of $\nu$ in this range in Fig.~\ref{fig1}.
In the left panel, which corresponds to 
$\dot{C}=10^{-27}$~erg\,cm$^{-3}$\,s$^{-1}$ 
and $B_0=10^{-6}$~G, we obtain 
$g_{\rm max}=6.0 \times 10^{-16}$~GeV$^{-1}$ 
for the resonance mass $m_{\phi}=\omega_{p}$ with 
$r_\nu=10^{-13}$ [case (a)]. 
For $r_\nu=10^{-10}$ [case (b)] and $r_\nu=10^{-7}$ [case (c)], 
we have $g_{\rm max}=1.9 \times 10^{-14}$~GeV$^{-1}$ and 
$g_{\rm max}=6.0 \times 10^{-13}$~GeV$^{-1}$, respectively. 
As we observe in the left panel of Fig.~\ref{fig1}, the upper limits of 
$g_{\rm max}$ for these three cases are 
tighter than the observational bound derived from polarization measurements of 
the MWD stars. 
Away from the resonance axion mass, however, 
the values of $g_{\rm max}$ are much larger than those obtained at $m_{\phi}=\omega_p$.
For the choices of $\dot{C}$ and $B_0$ 
given in the left panel of Fig.~\ref{fig1}, 
the upper limits of $g_{\rm max}$ derived in 
the mass ranges $m_{\phi} \ll \omega_p$ and $m_{\phi} \gg \omega_p$ 
are much weaker than those computed at $m_{\phi}=\omega_p$.
This reflects the fact that the resonance width is given by $\Delta m_{\phi}=\nu \ll \omega_p$, 
in which case the resonance occurs in a narrow region around $m_{\phi}=\omega_p$.

If we were to obtain tighter observational bounds on $\dot{C}$, Eq.~(\ref{gmax}) shows 
that $g_{\rm max}$ gets smaller for any mass $m_{\phi}$. 
This property also holds for $B_0$, where 
the observational finding of interstellar media with $B_0$ 
larger than $10^{-6}$~G 
results in smaller values of $g_{\rm max}$. 
In the right panel of Fig.~\ref{fig1}, 
we show $g_{\rm max}$ versus $m_{\phi}$ for  
$\dot{C}=10^{-31}$\,erg\,cm$^{-3}$\,s$^{-1}$ 
and $B_0=1$~G. As we see in Eq.~(\ref{gmax}), the coupling 
$g_{\rm max}$ for given $m_{\phi}$ and $\nu$ is $10^{-8}$ times 
as small as the one in the left panel. 
As a result, even the mass range outside the resonance point can 
enter the parameter space constrained by the MWD measurements.  
This tendency is more significant for increasing values of $\nu$, 
even if $g_{\rm max}$ at $m_{\phi}=\omega_p$ gets larger. 
In case (c), for example, the mass range satisfying the bound 
$g \le 5.4 \times 10^{-12}$~GeV$^{-1}$ corresponds to 
$1.1 \times 10^{-14}~{\rm eV} \le m_{\phi} \le 9.0 \times 10^{-11}~{\rm eV}$. 
Thus, for smaller $\dot{C}$ and larger $B_0$, 
it is possible to obtain tighter constraints 
on $g_{\rm max}$ with wider parameter spaces 
of $m_{\phi}$ away from the resonance point 
even if the resonance band is narrow.

While the data of $\dot{C}$ shown in 
Table~\ref{tabinter} correspond to those of the typical interstellar media, 
there may be possibilities to find some 
data of particular regions in the Milky Way galaxy or dwarf galaxies which result in tighter limits on $\dot{C}$ with larger $B_0$. 
Upcoming observations such as Vera Rubin Observatory and future Subaru HSC 
(Hyper Suprime-Cam) observations
 are expected to find other 
gas-rich faint dwarf galaxies like Leo T, which may obtain 
the cooling rate smaller than the current limit.
We also note that the plasma frequency $\omega_p$ can vary for a wider range than $10^{-12}~{\rm eV}$.
Then, we may be able to obtain more stringent 
constraints on $g_{\rm max}$ with the wider region of $m_{\phi}$ in future observations.

\section{Conclusions}
\label{consec}

In this paper, we proposed a new method of probing the axion-photon coupling 
through the heating of plasmas in interstellar media. We assumed the presence of 
an interstellar magnetic field aligned in a particular direction, whose typical value 
is of order $B_0=10^{-6}$~G. 
We also considered a nonrelativistic thermalized plasma of electrons and 
heavy ions. The observed cooling rate of interstellar media $\dot{C}$ 
and the thermal temperature $T$ are given in Table~\ref{tabinter}. 
For the typical electron number density $n_e \simeq 10^{-3}~{\rm cm}^{-3}$, 
the plasma frequency $\omega_p$ and the friction constant $\nu$ 
can be computed by Eq.~(\ref{nu}), so that 
$\nu \ll \omega_p \simeq 10^{-12}~{\rm eV}$ for 
all of these interstellar media. Although the precise values of $B_0$ have been 
unknown for the data shown in Table~\ref{tabinter}, we performed forecast constraints
 on the axion-photon coupling constant $g$ in preparation for upcoming observations 
 of the interstellar plasma.

In Sec.~\ref{dmsec}, we derived solutions 
to the electric field ${\bm E}$ and 
the vector potential ${\bm A}$ by dealing 
with the axion as nonrelativistic dark matter that oscillates around the potential minimum. 
We exploited the approximation that spatial gradient terms in the axion and electromagnetic 
field equations of motion can be neglected relative to time-dependent terms. In the 
$x$-direction parallel to the external 
magnetic field ${\bm B}_0$, we obtained analytic solutions to the $x$ components 
of ${\bm E}$ and ${\bm A}$.
We showed that both $E_x$ and $A_x$ are 
subject to forced resonance, with their 
leading-order solutions (\ref{Exf}) and 
(\ref{Axf}), respectively. 
In particular, the amplitude ${\cal D}$ 
in $A_x$ has a sharp peak around the 
resonance axion mass $m_{\phi}=\omega_p$.
We also found that the axion-photon 
coupling does not affect the dynamics 
in the $(y,z)$ plane.

In Sec.~\ref{heatsec}, we showed that the 
axion-photon coupling leads to the heating 
of interstellar media through the resonant 
energy conversion from axions to photons. 
In the direction parallel to the external 
magnetic field, the heating rate 
$\dot{Q}$ balances the energy damping rate $\dot{Q}_{\nu}$, 
which represents the energy release from photons to the plasma induced by the friction $\nu$. 
Since $\dot{Q}$ should not exceed the observed 
cooling rates $\dot{C}$ of interstellar media, 
it is possible to put upper bounds on  
the axion-photon coupling $g$. 
At the resonance point, $g$ is 
constrained as Eq.~(\ref{gcon}). 
For the typical interstellar magnetic field 
strength $B_0=10^{-6}$~G, we found that 
$g_{\rm max}$ at the resonance point 
($m_{\phi}=\omega_p$) can be more stringent than 
the limit constrained by the MWD measurements.  
As we see in the left panel of 
Fig.~\ref{fig1}, which corresponds to 
$B_0=10^{-6}$~G and $\dot{C}=10^{-27}$~erg\,cm$^{-3}$\,s$^{-1}$, $g_{\rm max}$ 
is much larger than the MWD bound for $m_{\phi}$ away from $\omega_p$. 
However, for larger $B_0$ and smaller $\dot{C}$, 
even the mass range outside the resonance point can enter the region constrained by the MWD 
measurements. This behavior is clearly seen 
in the right panel of Fig.~\ref{fig1}.

We have thus shown that observations of 
the interstellar magnetic field along with 
the cooling rates of plasmas will offer an interesting possibility for constraining 
the axion-photon coupling $g$. 
Since the resonance is most efficient for the mass $m_{\phi}$ around $\omega_p$, 
we may be able to obtain new bounds 
on $g$ for the mass region like
$10^{-15}~{\rm eV} \lesssim m_{\phi} \lesssim 
10^{-9}~{\rm eV}$ with upcoming data. 
This will bring a new perspective for 
scrutinizing the properties 
of light axions as a candidate for dark matter.

\section*{Acknowledgements}

We are grateful Shoichi Yamada for fruitful discussions. 
We also thank Zihui Wang for useful correspondence.
This work was supported by the Grant-in-Aid for Scientific Research Fund of 
the JSPS Nos.~20H05854 (TF), 22K03642 (ST), 23K03424 (TF),  and 
Waseda University Special Research Project No.~2024C-474 (ST). 

\appendix

\section*{Appendix I: Backreaction 
of electromagnetic fields}
\label{AppI}

We estimate the typical amplitude of the electric field ${\bm E}$ in the thermal plasma with temperature $T$. We ignore the contribution of the background magnetic field ${\bm B}$ for the moment and exploit the electric-field solution (\ref{Eso0}) for the time scale 
$t \lesssim 1/\nu$ (during which ${\bm E}$ is 
not damped by the friction $\nu$).
So long as $\omega_p \gg \nu$, we have 
$\dot{{\bm J}} \simeq \omega_p^2 {\bm E} 
\simeq \omega_p^2  {\bm E}_0 \cos (\omega_p t+\alpha_0)$ from Eq.~(\ref{Jeq}). 
This is integrated to give 
${\bm J} \simeq \omega_p 
{\bm E}_0 \sin (\omega_p t+\alpha_0)$, 
so that the averaged value of $|{\bm J}|^2$
over the oscillating period $2\pi/\omega_p$ yields
$|{\bm J}|^2 \simeq \omega_p^2 E_0^2/2$, 
where $E_0=|{\bm E}_0|$.
Since ${\bm J}=-n_e e {\bm u}_e$, 
it follows that 
\be
E_0^2 \simeq \frac{2n_e^2 e^2 |{\bm u}_e|^2}
{\omega_p^2}=2n_e m_e |{\bm u}_e|^2\,.
\ee
Using the typical thermal electron velocity
$|{\bm u}_e|=\sqrt{2T/m_e}$, we have 
\be
E_0 \simeq 2 \sqrt{n_e T}
\simeq 5.1 \times 10^{-9}\,{\rm eV}^2
\left( \frac{n_e}{10^{-3}\,{\rm cm}^{-3}} \right)^{1/2}
\left( \frac{T}{10^4\,{\rm K}} \right)^{1/2}\,.
\label{E0}
\ee
The background magnetic field of order 
$10^{-6}$\,G corresponds to 
$B_0=1.95 \times 10^{-8}$\,eV$^2$.
For $n_e \approx 10^{-3}\,{\rm cm}^{-3}$ 
and $T \approx 10^4$\,K, 
$E_0$ is of a similar order to  
$B_0 \approx 10^{-6}$\,G.

Let us confirm the validity of the approximation (\ref{apro1}). Since the axion amplitude $\bar{\phi}_0$ is related to the dark matter density as $\rho_{\rm DM}=m_{\phi}^2 \bar{\phi}_0^2/2$, the condition $g E_0 B_0 \ll m_{\phi}^2 \bar{\phi}_0$ translates to 
\be
g \ll 2.1 \times 10^{10}\,{\rm GeV}^{-1} 
\left( \frac{m_{\phi}}{10^{-12}\,{\rm eV}} \right)
\left( \frac{B_0}{10^{-6}\,{\rm G}} \right)^{-1}
\left( \frac{\rho_{\rm DM}}{0.3~{\rm GeV\,cm}^{-3}} \right)^{1/2}
\left( \frac{n_e}{10^{-3}~{\rm cm}^{-3}} \right)^{-1/2}
\left( \frac{T}{10^4~{\rm K}} \right)^{-1/2}\,.
\label{gos}
\ee
For the coupling $g$ constrained in the range (\ref{gcon}), 
the inequality (\ref{gos}) is well satisfied. 
This is also the case for $g_{\rm max}$ plotted 
in Fig.~\ref{fig1}. Thus, we can trust the homogenous solution 
(\ref{phiba}) of the axion field used in our analysis.

\section*{Appendix II: Solutions 
in the $(y,z)$ plane}
\label{AppII}

Here, we derive the solutions in the 
$(y,z)$ plane to 
Eqs.~\eqref{Ayeq}, \eqref{Azeq}, \eqref{Jyeq}, 
and \eqref{Jzeq}. 
For this purpose, we will consider the case 
in which the plasma frequency
is much larger than the cyclotron frequency, 
i.e., $\omega_p \gg \omega_c$.
On using the approximations $|k^2 A_y| \ll |\ddot{A}_y|$ and 
$|k^2 A_z| \ll |\ddot{A}_z|$ under 
the condition (\ref{kcon}), 
we have $J_y \simeq \ddot{A}_y \simeq -\dot{E}_y$ and 
$J_z \simeq \ddot{A}_z \simeq -\dot{E}_z$ from 
Eqs.~(\ref{Ayeq}) and (\ref{Azeq}). 
Substituting these relations into Eqs.~(\ref{Jyeq}) and (\ref{Jzeq}), 
it follows that 
\ba
& &
\ddot{E}_y+\nu \dot{E}_y+\omega_p^2 E_y \simeq 
-\omega_c \dot{E}_z\,,\label{Eyeq}\\
& &
\ddot{E}_z+\nu \dot{E}_z+\omega_p^2 E_z 
\simeq \omega_c \dot{E}_y\,.\label{Ezeq}
\ea
In comparison to Eq.~(\ref{Eeq0}), 
the presence of the magnetic field ${\bm B}_0$ 
gives rise to terms proportional to $\omega_c$ associated with the quasi-circular 
motion in the $(y,z)$ plane. 
Combining Eq.~(\ref{Eyeq}) with 
Eq.~(\ref{Ezeq}), we obtain 
the following fourth-order 
differential equations
\be
\ddddot{E_i}+2\nu \dddot{E_i}+\left( 2\omega_p^2+\omega_c^2+\nu^2 
\right) \ddot{E_i}+2\nu \omega_p^2 \dot{E_i}+\omega_p^4 E_i=0\,,
\ee
for both $i=y, z$. Assuming the solutions to these equations in the form $E_i=E_{i0}e^{\lambda t}$, where $E_{i0}$ and 
$\lambda$ are constants, we obtain the 
algebraic equation 
\be
\left( \lambda^2+\nu \lambda+\omega_p^2 \right)^2
+\omega_c^2 \lambda^2=0\,.
\label{lameq}
\ee
For $\omega_c=0$, the solution is given by  
$\lambda=(-\nu \pm \sqrt{\nu^2-4\omega_p^2})/2$, 
which reduces to 
$\lambda \simeq -\nu/2 \pm i \omega_p$ 
under the approximation $\omega_p \gg \nu$. 
For $\omega_c \neq 0$ with 
$\omega_p \gg \omega_c$, 
we assume the solution to Eq.~(\ref{lameq}) 
in the form 
$\lambda = -\nu/2 \pm i \omega_p+\epsilon$, 
where $|\epsilon| \ll \omega_p$. 
Substituting this into 
Eq.~(\ref{lameq}) and picking up the dominant contribution 
to $\epsilon$, it follows that 
\be
\lambda \simeq -\frac{\nu}{2} \pm i \omega_p
\pm \frac{i}{2} \omega_c\,,
\label{lamap}
\ee
which is valid for $\omega_p \gg \nu$ and $\omega_p \gg \omega_c$. 
Depending on the different signs in 
Eq.~(\ref{lamap}), we have 
four independent solutions to Eq.~(\ref{lameq}). 
Thus, the general solution to $E_y$ can 
be expressed as 
\be
E_y \simeq e^{-\nu t/2} \left[ c_1 \cos \left\{ \left( \omega_p+\frac{\omega_c}{2} 
\right)t+\alpha_1 \right\}+c_2 \cos \left\{ \left( \omega_p-\frac{\omega_c}{2} 
\right)t+\alpha_2 \right\} \right]\,,
\label{Ey}
\ee
where $c_1, c_2, \alpha_1, \alpha_2$ 
are constants. From Eq.~(\ref{Eyeq}), the other 
electric-field component $E_z$ 
is related to $E_y$. 
Dropping the contribution to $E_z$ from $\nu$ except the damping 
factor $e^{-\nu t/2}$ and using the approximation $\omega_p \gg \omega_c \neq 0$, 
we find 
\be
E_z \simeq e^{-\nu t/2} \left[ c_1 \left( 1-\frac{\omega_c}{4\omega_p} \right)
\sin \left\{ \left( \omega_p+\frac{\omega_c}{2} 
\right)t+\alpha_1 \right\}-c_2 \left( 1+\frac{\omega_c}{4\omega_p} \right)
\sin \left\{ \left( \omega_p-\frac{\omega_c}{2} 
\right)t+\alpha_2 \right\} \right]\,.
\label{Ez}
\ee
In both $E_y$ and $E_z$ components, there are two oscillating modes characterized by 
the frequencies $\omega_p \pm \omega_c/2$. 
So long as $\omega_p \gg \omega_c$, they are dominated by the plasma frequency with the oscillating time scale 
$t_p \approx 1/\omega_p$. 
For $\omega_p \gg \nu$, $t_p$ 
is much shorter than the damping time scale 
$t_d \approx 1/\nu$ induced by the 
friction term. 
When $\omega_c \neq 0$, the phases of oscillating terms in $E_y$ and $E_z$ 
are correlated with each other. 
If the initial conditions of $E_y$ and $E_z$ are chosen to be $c_2=0$, 
for example, we have $E_y^2/a^2+E_z^2/b^2=1$, where 
$a=c_1 e^{-\lambda t/2}$ and $b=c_1[1-\omega_c/(4\omega_p)] e^{-\lambda t/2}$.
This corresponds to the ellipse where the major radius $a$ and the minor 
radius $b$ decrease with the time scale $t_d$. 
As we observe in Eqs.~(\ref{Ayeq})-(\ref{Azeq}) and (\ref{Jyeq})-(\ref{Jzeq}), 
the coupling between axions and photons does 
not appear in the dynamics in the ($y,z$) 
plane and hence the electric field component $E_y$ or $E_z$ is not enhanced. 
While the above solutions to $E_y$ and $E_z$ 
have been derived under the condition 
$\omega_p \gg \omega_c$, the absence of 
energy transfer from axions to photons 
in the ($y,z$) plane persists for general 
values of $\omega_p$ and $\omega_c$.

\bibliographystyle{mybibstyle}
\bibliography{bib}

\end{document}